\documentclass[onecolumn]{article}
\usepackage{amsfonts}

\newtheorem{theorem}{Theorem}

\newtheorem{conjecture}[theorem]{Conjecture}

\newtheorem{lemma}[theorem]{Lemma}

\newenvironment{proof}[1][Proof]{\noindent\textbf{#1.} }{\ \rule{0.5em}{0.5em}}
\input{tcilatex}

\begin{document}

\title{Can entanglement efficiently be weakened by symmetrization?}
\author{Keiji Matsumoto \\
Quantum Information Science group, National Institute of Informatics,\\
2-1-2 Hitotsubashi, Chiyoda-ku, Tokyo 101-8430}
\maketitle

\begin{abstract}
Consider a quantum system with $m$ subsystems with $n$ qubits each, and
suppose the state of the system is living in the symmetric subspace. It is
known that, in the limit of $m\to\infty$, entanglement between any two
subsystems vanishes.

In this paper we study asymptotic behavior of the entanglement, or the
minimum trace distance from the totality of separable staes, as $m$ and $n$
grows. Our conjecture is that if $m$ is a polynomially bounded function in $%
n $, then the entanglement decreases polynomially.

The motivation of this study is a study of quantum Merlin-Arthur game. If
this conjecture is ture, we can prove that bipartite separable certificate
does not increase the computational power of the proof system. protocol.

In the paper, we provide two evidences which support the conjecture. The
first one shows that the entanglement is weakened polynomially fast as the
number $m$ of subsystems grows. Second evidence suggests that $m$ may be
polynomial in the number of qubits in each subsystems.
\end{abstract}

\section{Introduction}

In construction of a protocol for an interactive proof system or similar
proof systems~\cite{Kit99, Watrous:2005}, it is often a key whether one can
assume the separability between some quantum systems. For example, suppose
one wants to check whether a given pair of pure states are identical with
each other or not. If the given pair is separable, we can test the property
by the projection onto symmetric subspace, which is realized by a simple
algorithm using controlled swap. Such algorithm, however, fails if the given
pair is not necessarily separable.

Kobayashi \textit{et.al.} \cite{KMY} proposed a language class $\mathrm{QMA}%
(2)$, which is accepted by a quantum Merlin-Arthur (q.m.a., hereafter) game
with bipartite quantum certificate. By definition, $\mathrm{QMA}(2)\supset%
\mathrm{QMA}$. However, it is not trivial whether $\mathrm{QMA}(2)\subset%
\mathrm{QMA}$. Obviously, if one could check the separability of the
certificate, $\mathrm{QMA}(2)=\mathrm{QMA}$. However, such test is
impossible, for the totality of separable state and entangled states cannot
be separated by a hyperplane in the state space.(Here, note that we are
given only single copy of the quantum certificate). Therefore, the relation
between these computational class is not trivial.

Watrous \cite{Watrous:personal} had suggested the following q.m.a. protocol
which simulate q.m.a protocol with bipartite separable certificate with the
one with single certificate. In that, Merlin is asked to give $m$ copies of
the bipartite separable certificate. Arthur checks the symmetry and uses the
first part of the first copy and the second part of the second copy. If this
protocol is valid, $\mathrm{QMA}(2)=\mathrm{QMA}$.

For this protocol to be valid the reduced state has to be almost separable.
This is intuitively true, for the following reason. One party cannot have
entanglement with the rest of the system more than 
\[
\log \dim \mathbb{C}^{2^n}=n, 
\]
which should be equally distributed to each party. Hence, intuitively,
entanglement between two parties should be roughly 
\[
\frac{n}{m}, 
\]
which is small if $m$ is appropriately chosen polynomial function in $n$.

The purpose of the paper is to show two theorems which supports this
conjecture. The first one shows that the entanglement, or the minimum trace
distance from the totality of the separable state, is decreasing
polynomially fast as the number $m$ of subsystems grows. However, in the
proof, we have to set $m$ is an exponential function in $n$. On the other
hand, our second theorem shows that $m$ can be made polynomial for a certain
maximally entangled state.

The paper is organized as follows. First, we describe the conjectures and
its use in rigorous manner. Then, we proceed to prove two evidences which
supports our conjectures.

\section{Conjectures and its application}

\subsection{Conjectures}

Let us restate the problem rigorously. Let $\mathcal{H}^{\odot m}$\ denote a
symmetric subspace, or a subspace of $\mathcal{H}^{\otimes m}$ which is
spanned by vectors of the form $|\phi \rangle ^{\otimes m}$. Let $\mathcal{H}%
_1\simeq \mathcal{H}_2\simeq\cdots\simeq\mathbb{C}^{2^n}$. Given $\rho \in 
\mathcal{S}(\bigotimes_{i=1}^{m}\mathcal{H}_{i})$, we denote the reduced
density matrix to the composition of the first and the second system by $%
\left.\rho\right\vert _{\mathcal{H}_1\otimes \mathcal{H}_2}$. Also, define 
\[
d_{n,m}:=\dim ( \mathbb{C}^{2^n})^{\odot m}=\frac{(2^n+m-1)!}{(2^n-1)!m!}. 
\]
If \textit{one of} following conjectures is true, the protocol in \cite%
{Watrous:personal}, which will be fully stated in the next subsection,is
valid.

\begin{conjecture}
\label{conjecture1}Let $m$ be a appropriately chosen polynomial function in $%
n$, and $\rho \in \mathcal{S}((\mathbb{C}^{2^n})^{\odot m})$. Then, 
\begin{equation}
\min_{\sigma :\mathrm{separable}} \Vert\left.\rho \right\vert_{\mathcal{H}%
_1\otimes\mathcal{H}_2}-\sigma\Vert _1<\frac{1}{q}  \label{eqn:sym-separable}
\end{equation}
holds for a polynomial function $q$ in $n$.
\end{conjecture}

\begin{conjecture}
\label{conjecture2}Let $m$ be a appropriately chosen polynomial function in $%
n$, and suppose $\rho \in \mathcal{S}\left( (\mathbb{C}^{2^{n}})^{\odot 
\frac{m}{2}}\otimes (\mathbb{C}^{2^{n}})^{\odot \frac{m}{2}}\right) $, such
that the first $(\mathbb{C}^{2^{n}})^{\odot \frac{m}{2}}$ is a subset of $%
\bigotimes_{i=1}^{\frac{m}{2}}\mathcal{H}_{2i-1}$, and the second one is a
subset of $\bigotimes_{i=1}^{\frac{m}{2}}\mathcal{H}_{2i}$. Suppose also $%
\rho \in \mathcal{S}\left( (\mathbb{C}^{2^{n}}\otimes \mathbb{C}%
^{2^{n}})^{\odot \frac{m}{2}}\right) $, where $\mathbb{C}^{2^{n}}\otimes 
\mathbb{C}^{2^{n}}=\mathcal{H}_{2i-1}\otimes \mathcal{H}_{2i}$ . Then, (\ref%
{eqn:sym-separable}) holds for a polynomial function $q$ in $n$.
\end{conjecture}

Note that Conjecture \ref{conjecture1} is a stronger assertion than
Conjecture \ref{conjecture2}..

\subsection{How to utilize these properties}

The content of this section comes form \cite{Watrous:personal}.

Suppose that we can fabricate the projection onto symmetric subspace with
small error efficiently. (An example of quantum circuit achieving this is
explained later.) Using the projection measurement, we test the given state.
If the given state is in the symmetric subspace, the test accepts it with
high probability. Otherwise, the test still accept it, sometimes with
non-negligible property. However, upon acceptance, the output state is very
close to a symmetric state.

Therefore, using the projection onto symmetric subspace, we can always
obtain $\rho \in \mathcal{S}\left( (\mathbb{C}^{2^{n}})^{\odot m}\right) $
upon acceptance, and  $\left. \rho \right\vert _{\mathcal{H}_{1}\otimes 
\mathcal{H}_{2}}$, is almost separable, if  Conjecture \ref{conjecture1} is
true.

Suppose Conjecture \ref{conjecture2} is true (recall Conjecture \ref%
{conjecture1} is a stronger assertion than this). First, we apply projection
onto $(\mathbb{C}^{2^{n}}\otimes \mathbb{C}^{2^{n}})^{\odot \frac{m}{2}}$.
Then, the resultant state, upon acceptance, is convex combination of the
pure states of the form 
\[
\sum_{\mathbf{n}}\sqrt{a_{\mathbf{n}}}|\phi _{\mathbf{n}}\rangle ,
\]%
where $|\phi _{\mathbf{n}}\rangle \in \mathcal{W}_{\mathbf{n,}odd}\otimes 
\mathcal{W}_{\mathbf{n,}even}$ and $\mathcal{W}_{\mathbf{n},odd}$ and $%
\mathcal{W}_{\mathbf{n},even}$ are the homogeneous component of the
representation of $S_{\frac{m}{2}}$ on $\bigotimes_{i=1}^{\frac{m}{2}}%
\mathcal{H}_{2i-1}$ and $\bigotimes_{i=1}^{\frac{m}{2}}\mathcal{H}_{2i}$,
respectively, corresponding to the Yung index $\mathbf{n}$. For the proof of
this fact, see the proof of Lemma 1 in \cite{MatsumotoHayashi}. Note that
the assertion of this lemma uses $m$-copies of the pure state is invariant
by the action of $S_{m}$. In our case, the given state is not $m$-copies of
the pure state, but yet satisfies symmetry, which is enough to lead to the
same assertion. Now, we apply the projection onto $(\mathbb{C}%
^{2^{n}})^{\odot \frac{m}{2}}$ to $\bigotimes_{i=1}^{\frac{m}{2}}\mathcal{H}%
_{2i-1}$. Then, upon acceptance, the state collapses to $|\phi _{\mathbf{n}%
^{\prime }}\rangle $, where $\mathbf{n}^{\prime }$ equals $(\frac{m}{2}%
,0,\cdots ,0)$ and corresponds to symmetric subspace. Therefore, due to
Conjecture \ref{conjecture2}, $\left. \rho \right\vert _{\mathcal{H}%
_{1}\otimes \mathcal{H}_{2}}$ is almost separable.

Such procedures can be used to study a language class called $\mathrm{QMA}(2)
$, which is accepted by a quantum Merlin-Arthur (q.m.a., in short) game with
bipartite separable quantum certificate \cite{KMY}. By definition, $\mathrm{%
QMA}(2)\supset \mathrm{QMA}$. To show $\mathrm{QMA}(2)\subset \mathrm{QMA}$,
we construct q.m.a. protocol with (not necessarily separable) certificate
which simulates given or a q.m.a. protocol with a bipartite separable
certificate. In stead of single pair of certificates, Merlin provides many
pair of certificate. Arthur use the protocol above to obtain a bipartite
nearly separable quantum certificate.

\subsection{Testing symmetry}

Here, we describe how to approximately implement a projection onto symmetric
subspace. Construction is very similar to the main algorithm of \cite%
{Watrous:2000}. Suppose there is a circuit which generates 
\[
\frac{1}{\sqrt{\left\vert S_{m}\right\vert }}\sum_{\pi \in S_{m}}|\pi
\rangle |\psi _{\pi }\rangle ,
\]%
where $\langle \pi |\pi ^{\prime }\rangle =\delta _{\pi \pi ^{\prime }}$.
Here, we also suppose that this circuit resets working space in the end.

Given an input $|\phi \rangle $, this gate is applied to $|0\rangle
|\phi\rangle $, to obtain 
\[
\frac{1}{\sqrt{\left\vert S_m\right\vert }}\sum_{\pi \in S_m}|\pi \rangle
|\psi _\pi\rangle |\phi \rangle . 
\]
Then, controlled$\pi $ gate is applied, letting the control being the first
register, and the target being the second register: 
\[
\frac{1}{\sqrt{\left\vert S_m \right\vert }}\sum_{\pi \in S_m} |\pi\rangle
|\psi _\pi \rangle \pi |\phi\rangle . 
\]
Finally, we apply projection onto $\frac{1}{\sqrt{\left\vert
S_{m}\right\vert }}\sum_{\pi \in S_{m}}|\pi \rangle |\psi _{\pi }\rangle $
to the first register. This projection is implemented by running the state
generation circuit backwards, and measuring output by the computational
basis.

The final state is 
\begin{eqnarray*}
\frac{1}{\sqrt{\left\vert S_m\right\vert }} \sum_{\pi ,\pi^{\prime }\in S_m}
\langle \pi^\prime|\pi\rangle\langle\psi_{\pi^\prime} |\psi_\pi\rangle\pi
|\phi\rangle \\
=\frac{1}{\sqrt{\left\vert S_m\right\vert}} \sum_{\pi \in S_m}\pi
|\phi\rangle,
\end{eqnarray*}
which is an element of the symmetric subspace. It is easy to see that the
final state equals the input $|\phi \rangle $ if and only if $|\phi \rangle$
is an element of a symmetric subspace. Therefore, our circuit satisfies the
requirement.

Now, it still remains to show the construction of circuits which outputs $%
\frac{1}{\sqrt{\left\vert S_m\right\vert }}\sum_{\pi\in S_m}|\pi\rangle
|\psi _\pi\rangle $. We are done if there is a classical algorithm which
generates each $\pi$ uniformly randomly. This is easy, for each element of $%
S_n$ corresponds to an ordering of $1,2,\cdots ,n$.

\section{Evidences for the conjectures}

\subsection{Exponentially many sites}

In this section, we prove that the minimum trace distance from the totality
of separable states decreases polynomially fast as $m$ increases.

For that, we use entanglement of formation (EoF, in short), denoted by $%
E_{f}(\rho )$. This is an important measure of entanglement, first proposed
in \cite{BDSW}. Given state $\rho \in \mathcal{S}\left(\mathcal{H}_1\otimes 
\mathcal{H}_2\right)$, it is defined by 
\[
E_f(\rho ) :=\inf_{\left\{ p_i,\left\vert \phi_i\right\rangle \right\}}
\sum_{i}p_i S\left(\left. \left\vert \phi_i\right\rangle
\left\langle\phi_i\right\vert \right\vert_{\mathcal{H}_1}\right) , 
\]
where $S(\rho ):=-\mathrm{tr}\rho\log\rho$ and $\inf$ is taken over all the
pure state ensemble $\{p_i, |\phi_i\rangle \}$ with $\sum_ip_i|\phi_i\rangle%
\langle\phi_i| =\rho $. The following lemma is of interest in its own right,
due to importance of EoF.

\begin{lemma}
\label{lem:eof-sep}Let $m$ be a appropriately chosen \textrm{exponential}
function in $n$, and $\rho \in \mathcal{S}((\mathbb{C}^{2^n})^{\odot m})$.
Then, there is an \textrm{exponential} function $q$ which satisfies%
\[
E_{f}\left( \left.\rho\right\vert_{\mathcal{H}_1\otimes \mathcal{H}%
_2}\right) \leq \frac{1}{q} 
\]
\end{lemma}

\begin{proof}
It suffices to show the assertion when $\rho $ is pure. Let $\mu $ be a Haar
measure in $\mathrm{SU}(\mathbb{C}^{2^{n}})$ with normalization $\int \mu (%
\mathrm{d}U)=1$. Observe 
\begin{eqnarray*}
\left. \rho \right\vert _{\bigotimes_{i=3}^{m}\mathcal{H}_{i}} &\in &%
\mathcal{S}\left( (\mathbb{C}^{2^{n}})^{\odot (m-2)}\right) , \\
\left. \rho \right\vert _{\mathcal{H}_{1}\otimes \mathcal{H}_{2}} &\in &%
\mathcal{S}\left( (\mathbb{C}^{2^{n}})^{\odot 2}\right) 
\end{eqnarray*}%
and 
\[
I_{(\mathbb{C}^{2^{n}})^{\odot m}}=d_{n,m}\int (U|0^{n}\rangle \langle
0^{n}|U^{\dagger })^{\otimes m}\mu (\mathrm{d}U).
\]%
Using these facts, we have%
\begin{eqnarray*}
&&\left. \rho \right\vert _{\mathcal{H}_{1}\otimes \mathcal{H}_{2}} \\
&=&d_{n,m}\int (\langle 0^{n}|U^{\dagger })^{\otimes m-2}\rho
(U|0^{n}\rangle )^{\otimes m-2}\mu (\mathrm{d}U). \\
&=&d_{n,m}\int p_{U}\frac{(\langle 0^{n}|U^{\dagger })^{\otimes m-2}\rho
(U|0^{n}\rangle )^{\otimes m-2}}{p_{U}}\mu (\mathrm{d}U),
\end{eqnarray*}%
where $p_{U}=\mathrm{Tr}\left( \langle 0^{n}|U^{\dagger }\right) ^{\otimes
m-2}\rho \left( U|0^{n}\rangle \right) ^{\otimes m-2}$. Note the last end of
equation gives a decomposition of $\left. \rho \right\vert _{\mathcal{H}%
_{1}\otimes \mathcal{H}_{2}}$ into pure states. ($(\langle 0^{n}|U^{\dagger
})^{\otimes m-2}\rho (U|0^{n}\rangle )^{\otimes m-2}$ is of rank $1$ because 
$\rho $ is a pure state by assumption. ) Let $a_{U,i}$ ($a_{U,1}\geq
a_{U,2}\geq \cdots a_{U,d_{n,2}}$) be a Schmidt coefficient of the state $%
\frac{\left( \langle 0^{n}|U^{\dagger }\right) ^{\otimes m-2}\rho \left(
U|0^{n}\rangle \right) ^{\otimes m-2}}{p_{U}}$. Then we have%
\begin{eqnarray*}
&&d_{n,m}\int p_{U}a_{U,1}\mu (\mathrm{d}U) \\
&=&\hspace{-1mm}\max_{|\phi \rangle \in (\mathbb{C}^{2^{n}})^{\odot 2}}%
\hspace{-1mm}d_{n,m}\hspace{-1mm}\int \hspace{-2mm}p_{U}\langle \phi |\frac{%
\left( \langle 0^{n}|U^{\dagger }\right) ^{\otimes m-2}\rho \left(
U|0^{n}\rangle \right) ^{\otimes m-2}}{p_{U}}|\phi \rangle \mu (\mathrm{d}U)
\\
&\geq &\hspace{-2mm}d_{n,m}\hspace{-1mm}\int \hspace{-1mm}p_{U}\hspace{-1mm}%
\left( \langle 0^{n}|U^{\dagger }\right) ^{\otimes 2}\hspace{-1mm}\frac{%
\left( \langle 0^{n}|U^{\dagger }\right) ^{\otimes m-2}\hspace{-1mm}\rho 
\hspace{-1mm}\left( U|0^{n}\rangle \right) ^{\otimes m-2}}{p_{U}}\hspace{-1mm%
}\left( U|0^{n}\rangle \right) ^{\otimes 2}\hspace{-1mm}\mu (\mathrm{d}U) \\
&=&d_{n,m}\int \mathrm{Tr}\rho \left( U|0^{n}\rangle \langle
0^{n}|U^{\dagger }\right) ^{\otimes m-2}\mu (\mathrm{d}U) \\
&=&\frac{d_{n,m-2}}{d_{n,m}}\mathrm{Tr}\rho  \\
&=&\frac{(2^{n}+m-3)!}{(2^{n})!(m-3)!}\frac{(2^{n})!(m-1)!}{(2^{n}+m-1)!} \\
&=&\frac{(m-1)(m-2)}{(2^{n}+m-3)(2^{n}+m-2)}
\end{eqnarray*}%
Here, letting $m=2^{n}$, 
\begin{eqnarray*}
&=&\frac{(2^{2n}-1)(2^{2n}-2)}{(2^{n}+2^{2n}-3)(2^{n}+2^{2n}-2)} \\
&\geq &1-\frac{c}{2^{n}}.
\end{eqnarray*}%
This implies the Schmidt coefficient sharply concentrates to the first
element $a_{U,i}$. Intuitively, this implies entanglement of $\left. \rho
\right\vert _{\mathcal{H}_{1}\otimes \mathcal{H}_{2}}$ is very small, for
the state is decomposed into states which is close to a separable state in
average. To prove our lemma, we have to upperbound $E_{f}$ using this
quantity. Observe that 
\begin{eqnarray*}
&&\sum_{i}(-a_{U,i}\log a_{U,i}) \\
&\leq &(-a_{U,1}\log a_{U,1}) \\
&&-\dim (\mathbb{C}^{2^{n}})^{\odot 2}\frac{1-a_{U,1}}{\dim (\mathbb{C}%
^{2^{n}})^{\odot 2}}\log \frac{1-a_{U,1}}{\dim (\mathbb{C}^{2^{n}})^{\odot 2}%
} \\
&=&h(a_{U,1})+(1-a_{U,1})\log \dim (\mathbb{C}^{2^{n}})^{\odot 2},
\end{eqnarray*}%
with $h(x)=-x\log x-(1-x)\log (1-x)$. Therefore, if $n$ is large enough, we
have, 
\begin{eqnarray*}
&&E_{f}\left( \left. \rho \right\vert _{\mathcal{H}_{1}\otimes \mathcal{H}%
_{2}}\right)  \\
&\leq &d_{n,m}\int p_{U}\sum_{i}(-a_{U,i}\log a_{U,i})\mu (\mathrm{d}U) \\
&\leq &d_{n,m}\int p_{U}h(a_{U,1})\mu (\mathrm{d}U) \\
&&+(\log (d_{n,2}-1))d_{m,n}\int p_{U}(1-a_{U,1})\mu (\mathrm{d}U) \\
&\leq &h\left( d_{n,m}\int p_{U}a_{U,1}\mu (\mathrm{d}U)\right)  \\
&&+\log \left\{ \frac{(2^{n}+1)!}{(2^{n})!}-1\right\} d_{n,m}\int
p_{U}(1-a_{U,1})\mu (\mathrm{d}U) \\
&\leq &h\left( 1-\frac{c}{2^{n}}\right) +n(\log 2)\frac{c}{2^{n}} \\
&=&-\left( 1-\frac{c}{2^{n}}\right) \log \left( 1-\frac{c}{2^{n}}\right)  \\
&&-\frac{(c\log c)n}{2^{n}}+n(\log 2)\frac{c}{2^{n}} \\
&=&O\left( n2^{-n}\right) ,
\end{eqnarray*}%
and the proof is complete. Here, the first inequality is due to the
definition of EoF. The second inequality is proved by the maximizing entropy
for given $a_{U,1}$. The maximum is achieved by setting $a_{U,i}=\frac{%
1-a_{u,1}}{d_{n,2}-1}$ for $i\geq 2$. The third inequality is due to
convexity of $h(\,\cdot \,)$.
\end{proof}

\begin{theorem}
Let $m$ be a appropriately chosen \textrm{exponential} function in $n$, and $%
\rho \in \mathcal{S}((\mathbb{C}^{2^n})^{\odot m})$. Then, there is an 
\textrm{exponential} function $q$ which satisfies (\ref{eqn:sym-separable}).
\end{theorem}

\begin{proof}
Recall the well-known inequality 
\[
E_{f}(\rho )\geq E_{R}(\rho ), 
\]%
where 
\[
E_{R}(\rho ):=\min_{\sigma :\mathrm{separable}}D(\rho ||\sigma ) 
\]%
is called relative entropy of entanglement, first defined in \cite{VPJK:1997}%
. Also, recall well-known quantum version of Pinsker's inequality 
\[
D(\rho ||\sigma )\geq \frac{1}{2}\left( \left\Vert \rho -\sigma
\right\Vert_1\right)^2. 
\]%
These inequalities lead to 
\[
\min_{\sigma :\mathrm{separable}}\left\Vert \rho -\sigma \right\Vert _1\leq 
\sqrt{2E_{f}(\rho )}, 
\]%
which, combined with Lemma \ref{lem:eof-sep}, implies the theorem.
\end{proof}

\subsection{A maximally entangled state}

Previous theorem may not be a good evidence for our conjecture, for $m$
might have to be exponentially large. In this section, we supply an evidence
that polynomially many copies may be enough: We prove Conjecture \ref%
{conjecture2} when $\rho $ is a symmetric maximally entangled state. (Here
entanglement is understood in terms of $\bigotimes_{i=1}^{\frac{m}{2}}%
\mathcal{H}_{2i}$ -$\bigotimes_{i=1}^{\frac{m}{2}}\mathcal{H}_{2i-1}$%
-partition): 
\begin{eqnarray*}
\rho &=&|\Phi \rangle \langle \Phi |, \\
|\Phi \rangle &=&\sqrt{d_{n,\frac{m}{2}}}\int \left( U|0^n\rangle \overline{U%
}|0^n\rangle \right) ^{\otimes \frac{m}{2}}\mu (\mathrm{d}U).
\end{eqnarray*}%
To see that $|\Phi \rangle $ is maximally entangled, see partial transpose
of it,%
\begin{eqnarray*}
|\Phi \rangle ^{PT} &=&\sqrt{d_{n,\frac{m}{2}}}\int \left( U|0^n\rangle
\langle 0^n|U^{\dagger }\right) ^{\otimes \frac{m}{2}}\mu (\mathrm{d}U) \\
&=&\frac{1}{\sqrt{d_{n,\frac{m}{2}}}}I_{(\mathbb{C}^{2^n})^{\odot \frac{m}{2}%
},}
\end{eqnarray*}%
where the second identity is due to Shur's lemma.

\begin{theorem}
Let $m$ be a appropriately chosen polynomial function in $n$. and let $\rho $
be as stated above. Then, (\ref{eqn:sym-separable}) holds for a polynomial
function $q$ in $n$.
\end{theorem}

\begin{proof}
\begin{eqnarray*}
&&\left. \rho \right\vert _{\mathcal{H}_{1}\otimes \mathcal{H}_{2}} \\
&=&d_{n,\frac{m}{2}}\int \mu (\mathrm{d}U)\mu (\mathrm{d}U^{\prime }) \\
&&\times U|0^{n}\rangle \overline{U}|0^{n}\rangle \langle 0^{n}|U^{\prime
\dagger }\langle 0^{n}|\overline{U^{\prime }}^{\dagger }|\langle
0^{n}|U^{\prime \dagger }U|0^{n}\rangle |^{m-2}.
\end{eqnarray*}%
Here we decompose%
\[
U^{\prime }|0^{n}\rangle =(\cos \theta )U|0^{n}\rangle +(\sin \theta )V|\phi
_{U}\rangle ,
\]%
using a state $|\phi _{U}\rangle $ and $V$ \ with $\langle \phi
_{U}|U|0^{n}\rangle =0$ and $\langle \phi _{U}|V^{\dagger }U|0^{n}\rangle =0$%
. Here, $\theta $ and $V$ runs all over the interval $[0,\pi )$, and $%
\mathrm{SU}(2^{n}-1)$. (Here, more rigorously, we are considering an
embedding of $\mathrm{SU}(2^{n}-1)$ into $\mathrm{SU}(2^{n})$ with $\langle
\phi _{U}|V^{\dagger }U|0^{n}\rangle =0$.) Also, 
\[
\mu (\mathrm{d}U^{\prime })=\nu _{n}(\mathrm{d}\theta )\,\mu ^{\prime }(%
\mathrm{d}V),
\]%
where $\mu ^{\prime }$ is the Haar measure in $\mathrm{SU}(2^{n}-1)$ with $%
\int \mu ^{\prime }(\mathrm{d}V)=1$, and $\nu _{n}$ is an appropriate
measure. Note also%
\[
\int V|\phi _{U}\rangle \mu ^{\prime }(\mathrm{d}V)=0.
\]%
Therefore, 
\begin{eqnarray*}
\left. \rho \right\vert _{\mathcal{H}_{1}\otimes \mathcal{H}_{2}} &=&d_{n,%
\frac{m}{2}}\int \left( \cos ^{m}\theta \right) \nu _{n}(\mathrm{d}\theta )%
\overline{\rho } \\
&&+d_{n,\frac{m}{2}}\int \left( \cos ^{m-2}\theta \sin ^{2}\theta \right)
\nu _{n}(\mathrm{d}\theta )X,
\end{eqnarray*}%
where 
\[
\overline{\rho }:=\int \,\mu (\mathrm{d}U)U|0^{n}\rangle \overline{U}%
|0^{n}\rangle \langle 0^{n}|U^{\dagger }\langle 0^{n}|\overline{U}^{\dagger
},
\]%
and 
\[
X:=\int \mu ^{\prime }(\mathrm{d}V)\mu (\mathrm{d}U)U|0^{n}\rangle \overline{%
U}|0^{n}\rangle \langle \phi _{U}|V^{\dagger }\langle \phi _{U}|\overline{V}%
^{\dagger }.
\]%
Note that $\overline{\rho }$ is a separable state, and $X$ satisfies $%
\mathrm{tr}X=0$. Note, letting $\cos \theta =|\langle \phi |U|0^{n}\rangle |$%
,%
\begin{eqnarray*}
\frac{1}{d_{n,\frac{m}{2}}} &=&\langle \phi |^{\otimes \frac{m}{2}}\int
\left( U|0^{n}\rangle \langle 0^{n}|U^{\dagger }\right) ^{\otimes \frac{m}{2}%
}\mu (\mathrm{d}U)|\phi \rangle ^{\otimes \frac{m}{2}} \\
&=&\int |\langle \phi |U|0^{n}\rangle |^{m}\nu _{n}(\mathrm{d}\theta )\mu
^{\prime }(\mathrm{d}V) \\
&=&\int \cos ^{m}\theta \nu _{n}(\mathrm{d}\theta ),
\end{eqnarray*}%
where the second equality is due to $\int \mu ^{\prime }(\mathrm{d}V)=1$.
Due to this,%
\begin{eqnarray*}
&&\int \sin ^{2}\theta \cos ^{m-2}\theta \,\,\nu _{n}(\mathrm{d}\theta ) \\
&=&\int \left( 1-\cos ^{2}\theta \right) \cos ^{m-2}\theta \,\,\nu _{n}(%
\mathrm{d}\theta ) \\
&=&\frac{1}{d_{n,\frac{m}{2}-1}}-\frac{1}{d_{n,\frac{m}{2}}} \\
&=&\frac{1}{d_{n,\frac{m}{2}}}\frac{2(2^{n}-1)}{m}.
\end{eqnarray*}%
Observe that%
\[
\left( 2^{n}-1\right) \int V|\phi _{U}\rangle \langle \phi _{U}|V^{\dagger
}\mu ^{\prime }(\mathrm{d}V)+U|0^{n}\rangle \langle 0^{n}|U^{\dagger }
\]%
is an identity operator in $%
\mathbb{C}
^{2^{n}}$. Therefore, its partial transpose with normalization factor
satisfies%
\begin{eqnarray*}
&&\frac{1}{\sqrt{2^{n}}}\left\{ 
\begin{array}{c}
\left( 2^{n}-1\right) \int V|\phi _{U}\rangle \overline{V}|\phi _{U}\rangle
\mu ^{\prime }(\mathrm{d}V) \\ 
+U|0^{n}\rangle \overline{U}|0^{n}\rangle 
\end{array}%
\right\}  \\
&=&\sqrt{2^{n}}\int U|0^{n}\rangle \overline{U}|0^{n}\rangle \mu (\mathrm{d}%
U) \\
&:&=|\Phi \rangle ,
\end{eqnarray*}%
or%
\begin{eqnarray*}
&&\int V|\phi _{U}\rangle \overline{V}|\phi _{U}\rangle \mu ^{\prime }(%
\mathrm{d}V) \\
&=&\frac{1}{2^{n}-1}\left( \sqrt{2^{n}}|\Phi \rangle -U|0^{n}\rangle 
\overline{U}|0^{n}\rangle \right) .
\end{eqnarray*}%
This leads to%
\begin{eqnarray*}
&&X=\frac{1}{2^{n}-1}\left( 
\begin{array}{c}
\sqrt{2^{n}}\int U|0^{n}\rangle \overline{U}|0^{n}\rangle \mu (\mathrm{d}%
U)\langle \Phi | \\ 
-\overline{\rho }%
\end{array}%
\right)  \\
&=&\frac{1}{2^{n}-1}\left( |\Phi \rangle \langle \Phi |-\overline{\rho }%
\right) .
\end{eqnarray*}%
Putting together,%
\begin{eqnarray*}
\left. \rho \right\vert _{\mathcal{H}_{1}\otimes \mathcal{H}_{2}} &=&%
\overline{\rho }+\frac{2(2^{n}-1)}{m}\frac{1}{2^{n}-1}\left( |\Phi \rangle
\langle \Phi |-\overline{\rho }\right)  \\
&=&\overline{\rho }+\frac{2}{m}\left( |\Phi \rangle \langle \Phi |-\overline{%
\rho }\right) .
\end{eqnarray*}%
For $\overline{\rho }$ is separable, $\left\Vert \frac{2}{m}\left( |\Phi
\rangle \langle \Phi |-\overline{\rho }\right) \right\Vert _{tr}\leq \frac{4%
}{m}$ implies our assertion.
\end{proof}

\section{Discussion}

We had presented two evidences for our conjecture. The first one shows that
the entanglement is weakened polynomially fast as the number $m$ of
subsystems grows. Second evidence suggests that $m$ may be polynomial in the
number of qubits in each subsystems.

Lemma \ref{lem:eof-sep}, used to show the first evidence, is of interest in
its own right, for EoF is a very important entanglement measure.

\section*{Acknowledgment}

The author thank J. Watrous and H. Kobayashi for discussions.


\begin{thebibliography}{9}
\bibitem{BDSW} C. H. Bennett, D. P. DiVincenzo, J. A. Smolin, W. K.
Wootters, \textquotedblleft Mixed--state entanglement and quantum error
correction\textquotedblright , Phys. Rev. A, vol. 54, no. 5, pp. 3824--3851,
1996.

\bibitem{Kit99} A. Kitaev. Quantum NP. Talk at the 2nd Workshop on
Algorithms in Quantum Information Processing, DePaul University, Chicago,
January 1999.

\bibitem{KMY} H. Kobayashi, K. Matsumoto, T. Yamakami, Quantum Merlin-Arthur
Proof Systems,: Are MultipleMerlins More Helpful to Arthur?, ISAAC 2003:
189-198.

\bibitem{MatsumotoHayashi} K. Matsumoto, H. Hayashi, Universal entanglement
concentration, quant-ph/0509140.

\bibitem{VPJK:1997} V. Vedral, M.B. Plenio, K. Jacobs, P. L. Knight,
Statistical Inference, Distinguishability of Quantum States, And Quantum
Entanglement, Phys.Rev. A56 4452, 1997.

\bibitem{Watrous:2000} J. Watrous. Succinct quantum proofs for properties of
finite groups. Proceedings of the 41st Annual Symposium on Foundations of
Computer Science, pages 537--546, 2000.

\bibitem{Watrous:personal} J. Watrous, personal communication, 2002.

\bibitem{Watrous:2005} C. Marriott and J. Watrous. Quantum Arthur-Merlin
games. Computational Complexity, 14(2): 122--152, 2005.
\end{thebibliography}
\end{document}